\documentclass[twocolumn,showpacs,preprintnumbers,amsmath,aps]{revtex4}
\usepackage{graphicx}
\usepackage{dcolumn}
\usepackage{bm}
\usepackage{multirow}
\usepackage{tabularx}
\usepackage{color}
\begin{document}
%%%%%%%%%%%%%%%%%%%%%%%%%%%%
\newcommand{\kvec}{\mbox{{\scriptsize {\bf k}}}}
%%%%%%%%%%%%%%%%%%%%%%%%%%%%
\def\eq#1{(\ref{#1})}
\def\fig#1{Fig.\hspace{1mm}\ref{#1}}
\def\tab#1{table\hspace{1mm}\ref{#1}}
%
%%%%%%%%%%%%%%%%%%%%%%%%%%%
%%%%%%%%%%%%%%%%%%%%%%%%%%%
%
\title{
---------------------------------------------------------------------------------------------------------------\\
Study of the superconducting state in the \emph{Cmmm} phase of ${\rm GeH_{4}}$ compound}
\author{R. Szcz{\c{e}}{\'s}niak$^{\left(1\right)}$}\email{szczesni@wip.pcz.pl} 
\author{A.P. Durajski$^{\left(1\right)}$}
\author{D. Szcz{\c{e}}{\'s}niak$^{\left(2, 3\right)}$}
%%%%%%%%%%%%
%\email{szczesni@wip.pcz.pl}
\affiliation{1. Institute of Physics, Cz{\c{e}}stochowa University of Technology, Ave. Armii Krajowej 19, 42-200 Cz{\c{e}}stochowa, Poland}
\affiliation{2. Institute of Physics, Jan D{\l}ugosz University in Cz{\c{e}}stochowa, Ave. Armii Krajowej 13/15, 42-200 Cz{\c{e}}stochowa, Poland}
\affiliation{3. Institute for Molecules and Materials UMR 6283, University of Maine, Ave. Olivier Messiaen, 72085 Le Mans, France}
%%%%%%%%%%%%
%%%%%%%%%%%%
\date{\today} 
\begin{abstract}
%%%%%%%%%%%%%%%%%%%%%%%%%%%%%%%%%%%%%%%%%%%%%
The superconducting state, that gets induced in the ${\rm GeH_{4}}$ compound under the pressure at $20$ GPa, is characterized by the high critical temperature $T_{C}\left(\mu^{\star}\right)\in\left<54.5,31.2\right>$ K, where $\mu^{\star}$ is the Coulomb pseudopotential, and $\mu^{\star}\in\left<0.1,0.3\right>$. Other thermodynamic parameters for the considered range of the Coulomb pseudopotential significantly differ from the predictions of the BCS model. In particular: (i) the ratio of the energy gap to the critical temperature ($R_{\Delta}\equiv 2\Delta\left(0\right)/k_{B}T_{C}$) changes from $4.07$ to $3.88$; (ii) the value of the parameter $R_{C}\equiv\Delta C\left(T_{C}\right)/C^{N}\left(T_{C}\right)$ decreases from $1.85$ to $1.69$, where $\Delta C$ denotes the specific heat jump, and $C^{N}$ is the specific heat of the normal state; (iii) the ratio $R_{H}\equiv T_{C}C^{N}\left(T_{C}\right)/H^{2}_{C}\left(0\right)$ increases from $0.147$ to $0.155$; $H_{C}$ represents the thermodynamic critical field. 
\end{abstract}
\maketitle

In the elements and the chemical compounds under the influence of high pressure ($p$), the electron-phonon interaction can be so strong that the superconducting state is characterized by a high value of the critical temperature. Of particular note are the results obtained for lithium and calcium, where the maximum value of the critical temperature is equal to: $\left[T_{C}\right]^{\rm Li}_{p=30.2{\rm GPa}}=14$ K and $\left[T_{C}\right]^{\rm Ca}_{p=161{\rm GPa}}=25$ K \cite{Deemyad}, \cite{Yabuuchi}. The other thermodynamic parameters of the superconducting state in Li and Ca, as the result of the existence of the retardation and strong-coupling effects, cannot be properly described in the framework of the BCS model \cite{SzczesniakA01}, \cite{SzczesniakA02}, \cite{SzczesniakA03}, \cite{BCS1}, \cite{BCS2}.

Probably, the highest value of the critical temperature can be observed in metallic hydrogen \cite{Ashcroft}. Unfortunately, also at an extremely high pressure. In particular, it has been predicted that hydrogen enters the metallic state at $400$ GPa \cite{Stadele}. It exists then in the molecular state, in which the superconducting state with the critical temperature of $240$ K has been predicted \cite{Cudazzo}. Note, that the superconducting state in the molecular hydrogen can be strongly anisotropic in some ranges of the pressure. In the simplest case, 
the three-band formalism should be used for the description of its properties \cite{Cudazzo}, \cite{SzczesniakA04}.

Above $500$ GPa, molecular hydrogen dissociates into a single atomic phase \cite{Yan}, \cite{McMahon}. In the examined case, the highest value of $T_{C}$ is predicted for the pressure at $2000$ GPa, where the critical temperature reaches $600$ K \cite{McMahon}. For the considered pressure, the low temperature energy gap at the Fermi surface ($2\Delta\left(0\right)$) is so large that the dimensionless ratio $2\Delta\left(0\right)/k_{B}T_{C}$ equals approximately $6$ \cite{SzczesniakA05}. The comparable value of $2\Delta\left(0\right)$ is only observed in the group of the high-temperature superconductors (cuprates) \cite{SzczesniakA06}, \cite{SzczesniakA07}.

In order to reduce the pressure of metallization in hydrogen, a method of the chemical precompression has been proposed \cite{Ashcroft01}. Physically, this idea can be realized in the group of the hydrogenated compounds of the following type: ${\rm SiH_{4}\left(H_{2}\right)_{2}}$, ${\rm Si_{2}H_{6}}$, ${\rm B_{2}H_{6}}$ and ${\rm PtH}$ \cite{Li}, \cite{Jin}, \cite{Kazutaka}, \cite{Kim}. 

Typically, the properties of the superconducting state for pressures that excess the value of $100$ GPa have been considered in the literature. Hence, it is difficult to verify the theoretical results in the experimental way. 

In the paper, we have determined the basic thermodynamic properties of the superconducting state in the \emph{Cmmm} phase of the ${\rm GeH_{4}}$ compound for the relatively low value of the pressure ($p=20$ GPa). In the examined case, the critical temperature, calculated on the basis of the modified McMillan formula, equals $\sim 47$ K \cite{Zhang}, \cite{McMillan}, \cite{AllenDynes}. Let us note that the low value of $p$ will allow in the future an easy comparison of the results obtained in this study with the experimental data.

Numerical calculations have been conducted in the framework of the Eliashberg formalism. On the imaginary axis the Eliashberg equations take the form \cite{Eliashberg1}, \cite{Eliashberg2}:

\begin{equation}
\label{r1}
\phi_{n}=\frac{\pi}{\beta}\sum_{m=-M}^{M}
\frac{\lambda\left(i\omega_{n}-i\omega_{m}\right)-\mu^{\star}\theta\left(\omega_{c}-|\omega_{m}|\right)}
{\sqrt{\omega_m^2Z^{2}_{m}+\phi^{2}_{m}}}\phi_{m},
\end{equation}
\begin{equation}
\label{r2}
Z_{n}=1+\frac{1}{\omega_{n}}\frac{\pi}{\beta}\sum_{m=-M}^{M}
\frac{\lambda\left(i\omega_{n}-i\omega_{m}\right)}{\sqrt{\omega_m^2Z^{2}_{m}+\phi^{2}_{m}}}
\omega_{m}Z_{m},
\end{equation}
where $\phi_{n}\equiv\phi\left(i\omega_{n}\right)$ denotes the order parameter function, and $Z_{n}\equiv Z\left(i\omega_{n}\right)$ is the wave function renormalization factor; $\omega_{n}$ stands for the Matsubara frequency: $\omega_{m}\equiv \left(\pi / \beta\right)\left(2m-1\right)$, where $\beta\equiv\left(k_{B}T\right)^{-1}$; $k_{B}$ is the Boltzmann constant. The order parameter is defined by the formula: $\Delta_{n}\equiv\phi_{n}/Z_{n}$. 

The pairing kernel for the electron-phonon interaction can be represented by:
$\lambda\left(z\right)\equiv 2\int_0^{\Omega_{\rm{max}}}d\Omega\frac{\Omega}{\Omega ^2-z^{2}}\alpha^{2}F\left(\Omega\right)$, where the Eliashberg function ($\alpha^{2}F\left(\Omega\right)$) has been determined in the paper \cite{Zhang}. The value of the maximum phonon frequency ($\Omega_{\rm max}$) is equal to $515.69$ meV. 

The symbol $\mu^{\star}$ represents the Coulomb pseudopotential, which models the depairing electron-electron interaction. The quantity $\theta$ denotes the Heaviside function, and $\omega_{c}$ is the cut-off frequency: $\omega_{c}=3\Omega_{\rm{max}}$.

The Eliashberg equations have been solved with the help of the numerical methods discussed and used in \cite{SzczesniakB01}. 
The convergence of the solutions has been obtained in the temperature range from $T_{0}=3$ K to $T_{C}$; $M=1100$ has been assumed.  

The form of the functions $\phi$ and $Z$ on the real axis ($\omega$) has been determined by using the Eliashberg equations in the mixed representation \cite{Marsiglio}:

\begin{widetext}
\begin{eqnarray}
\label{r3}
\phi\left(\omega+i\delta\right)&=&
                                  \frac{\pi}{\beta}\sum_{m=-M}^{M}
                                  \left[\lambda\left(\omega-i\omega_{m}\right)-\mu^{\star}\theta\left(\omega_{c}-|\omega_{m}|\right)\right]
                                  \frac{\phi_{m}}
                                  {\sqrt{\omega_m^2Z^{2}_{m}+\phi^{2}_{m}}}\\ \nonumber
                              &+& i\pi\int_{0}^{+\infty}d\omega^{'}\alpha^{2}F\left(\omega^{'}\right)
                                  \left[\left[N\left(\omega^{'}\right)+f\left(\omega^{'}-\omega\right)\right]
                                  \frac{\phi\left(\omega-\omega^{'}+i\delta\right)}
                                  {\sqrt{\left(\omega-\omega^{'}\right)^{2}Z^{2}\left(\omega-\omega^{'}+i\delta\right)
                                  -\phi^{2}\left(\omega-\omega^{'}+i\delta\right)}}\right]\\ \nonumber
                              &+& i\pi\int_{0}^{+\infty}d\omega^{'}\alpha^{2}F\left(\omega^{'}\right)
                                  \left[\left[N\left(\omega^{'}\right)+f\left(\omega^{'}+\omega\right)\right]
                                  \frac{\phi\left(\omega+\omega^{'}+i\delta\right)}
                                  {\sqrt{\left(\omega+\omega^{'}\right)^{2}Z^{2}\left(\omega+\omega^{'}+i\delta\right)
                                  -\phi^{2}\left(\omega+\omega^{'}+i\delta\right)}}\right],
\end{eqnarray}
and
\begin{eqnarray}
\label{r4}
Z\left(\omega+i\delta\right)&=&
                                  1+\frac{i}{\omega}\frac{\pi}{\beta}\sum_{m=-M}^{M}
                                  \lambda\left(\omega-i\omega_{m}\right)
                                  \frac{\omega_{m}Z_{m}}
                                  {\sqrt{\omega_m^2Z^{2}_{m}+\phi^{2}_{m}}}\\ \nonumber
                              &+&\frac{i\pi}{\omega}\int_{0}^{+\infty}d\omega^{'}\alpha^{2}F\left(\omega^{'}\right)
                                  \left[\left[N\left(\omega^{'}\right)+f\left(\omega^{'}-\omega\right)\right]
                                  \frac{\left(\omega-\omega^{'}\right)Z\left(\omega-\omega^{'}+i\delta\right)}
                                  {\sqrt{\left(\omega-\omega^{'}\right)^{2}Z^{2}\left(\omega-\omega^{'}+i\delta\right)
                                  -\phi^{2}\left(\omega-\omega^{'}+i\delta\right)}}\right]\\ \nonumber
                              &+&\frac{i\pi}{\omega}\int_{0}^{+\infty}d\omega^{'}\alpha^{2}F\left(\omega^{'}\right)
                                  \left[\left[N\left(\omega^{'}\right)+f\left(\omega^{'}+\omega\right)\right]
                                  \frac{\left(\omega+\omega^{'}\right)Z\left(\omega+\omega^{'}+i\delta\right)}
                                  {\sqrt{\left(\omega+\omega^{'}\right)^{2}Z^{2}\left(\omega+\omega^{'}+i\delta\right)
                                  -\phi^{2}\left(\omega+\omega^{'}+i\delta\right)}}\right], 
\end{eqnarray}
\end{widetext}
%%%%%%%%%%%%%%%%%%%%%%%%%%%
%
where the symbols $N\left(\omega\right)$ and $f\left(\omega\right)$ represent the Bose-Einstein and Fermi-Dirac functions, respectively. 

A way of solving the equations \eq{r3} and \eq{r4} has been discussed in \cite{SzczesniakB02}.

\begin{figure}[ht]
\includegraphics*[width=\columnwidth]{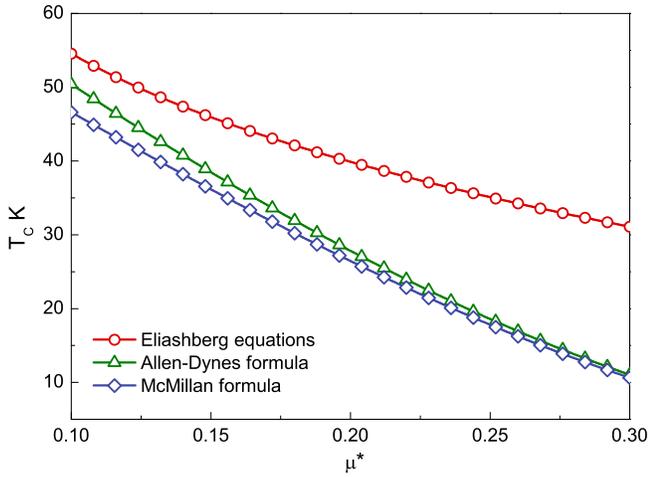}
\caption{The critical temperature as the function of the Coulomb pseudopotential; the results have been obtained by using the Eliashberg equations and the analytical formulas \cite{McMillan}, \cite{AllenDynes}.} 
\label{f1}
\end{figure}

\fig{f1} represents the dependence of the critical temperature on the Coulomb pseudopotential ($\mu^{\star}\in\left<0.1,0.3\right>$). The obtained results prove that $T_{C}$ is high in the whole range of $\mu^{\star}$ which has been considered. In particular: $T_{C}\in\left<54.5,31.2\right>$ K. Additionally in \fig{f1}, we have plotted the courses of the dependence of $T_{C}$ on $\mu^{\star}$ obtained with the help of the McMillan and Allen-Dynes expression \cite{McMillan}, \cite{AllenDynes}. It can be easily seen that the analytical formulas significantly underestimate the critical temperature, especially for the high values of the Coulomb pseudopotential.

%
%%%%%%%%%%%%%%%%%%%%%%%%%%%(Rysunek.1.)
%%%%%%%%%%%%%%%%%%%%%%%%%%%
%

%
%%%%%%%%%%%%%%%%%%%%%%%%%%%(Rysunek.2)
\begin{figure}[ht]
\includegraphics*[width=\columnwidth]{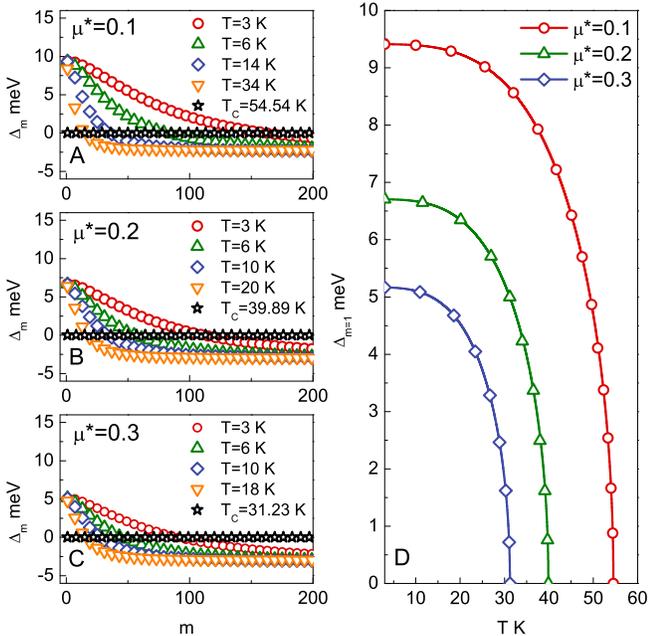}
\caption{(A)-(C) The form of the order parameter on the imaginary axis for the selected values of the temperature and the Coulomb pseudopotential. (D) The order parameter for $m = 1$ as the function of the temperature for the selected values of the Coulomb pseudopotential.}
\label{f2}
\end{figure}
%%%%%%%%%%%%%%%%%%%%%%%%%%%
%

In \fig{f2} (A)-(C), we have plotted the order parameter on the imaginary axis for the selected values of the temperature and the Coulomb pseudopotential. It has been found that with the increase in the number $m$, the function $\Delta_{m}$ strongly decreases, and then becomes saturated. The influence of the temperature and the Coulomb pseudopotential on the order parameter value can be the most conveniently traced on the example of the course of the function $\Delta_{m=1}\left(T,\mu^{\star}\right)$. It can be parameterized with the help of the formula: 
$\Delta_{m=1}\left(T,\mu^{\star}\right)=\Delta_{m=1}\left(T_{0},\mu^{\star}\right)\sqrt{1-\left(\frac{T}{T_{C}}\right)^{\beta}}$, 
where $\Delta_{m=1}\left(T_{0},\mu^{\star}\right)=58.4\left(\mu^{\star}\right)^{2}-44.58\mu^{\star}+13.286$ meV and $\beta=3.3$. 

It should be noted that the quantity $2\Delta_{m=1}\left(T,\mu^{\star}\right)$ with a good approximation reproduces the dependence of the energy gap at the Fermi surface on the temperature and the parameter $\mu^{\star}$. The obtained results have been shown in \fig{f2} (D). 

%
%%%%%%%%%%%%%%%%%%%%%%%%%%%(Rysunek.3.)
\begin{figure}[ht]
\includegraphics*[width=\columnwidth]{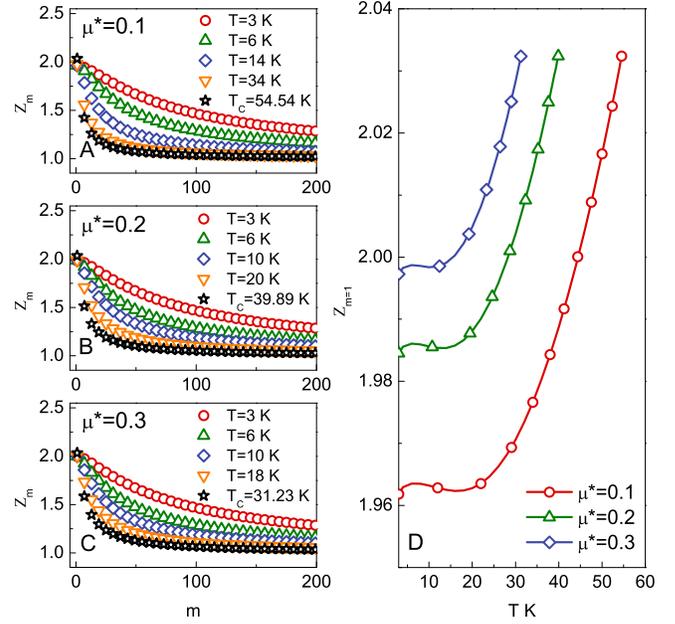}
\caption{(A)-(C) The form of the wave function renormalization factor on the imaginary axis for the selected values of the temperature and the Coulomb pseudopotential. (D) The renormalization factor for $m=1$ as the function of the temperature for the selected values of the Coulomb pseudopotential.} 
\label{f3}
\end{figure}
%%%%%%%%%%%%%%%%%%%%%%%%%%%
%

The second solution of the Eliashberg equations on the imaginary axis has been presented in \fig{f3} (A)-(C). Just as it was in the case of the order parameter, the value of $Z_{m}$ rapidly decreases with the increasing number $m$. On the other hand, the effect of the temperature and the Coulomb pseudopotential on the wave function renormalization factor is meager. 

It is worth noting that for $T=T_{C}$, the value of $Z_{m=1}$ is independent of the parameter $\mu^{\star}$. In the considered case, it can be calculated analytically using the formula: $Z_{m=1}=1+\lambda$, where the symbol $\lambda$ is the electron-phonon coupling constant: ${\lambda\equiv 2\int^{\Omega_{\rm{max}}}_0 \alpha^2\left(\Omega\right)F\left(\Omega\right)/\Omega}$. For the 
${\rm GeH_{4}}$ compound $\lambda$ is equal to $1.0324$, thus: $\left[Z_{m=1}\right]_{T=T_{C}}=2.0324$. The identical result has been obtained by using the numerical procedures, indicating a high accuracy of the computation carried out.

\begin{figure}[ht]
\includegraphics*[width=\columnwidth]{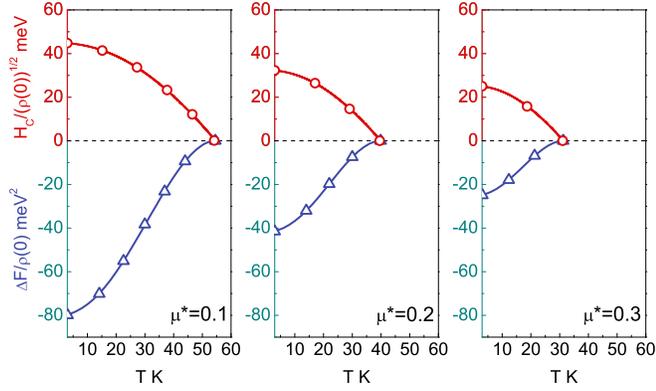}
\caption{The free energy difference (at the bottom) and the thermodynamic critical field (at the top) as the function of the temperature for the selected values of the Coulomb pseudopotential.} 
\label{f4}
\end{figure}

From the physical point of view, the wave function renormalization factor for $m=1$ with a very good approximation reproduces the value of the electron effective mass: $m^{\star}_{e}\simeq Z_{m=1}m_{e}$, where the symbol $m_{e}$ denotes the electron band mass. The results presented in \fig{f4} (D) prove that the electron effective mass in ${\rm GeH_{4}}$ is high in the entire range of existence of the superconducting state. 

The solutions of the Eliashberg equations on the imaginary axis have been used in the calculations of the free energy difference between the superconducting and the normal state \cite{Bardeen}: 
\begin{eqnarray}
\label{r5}
\frac{\Delta F}{\rho\left(0\right)}&=&-\frac{2\pi}{\beta}\sum_{n=1}^{M}
\left(\sqrt{\omega^{2}_{n}+\Delta^{2}_{n}}- \left|\omega_{n}\right|\right)\\ \nonumber
&\times&(Z^{S}_{n}-Z^{N}_{n}\frac{\left|\omega_{n}\right|}
{\sqrt{\omega^{2}_{n}+\Delta^{2}_{n}}}). 
\end{eqnarray}  
The symbols $Z^{S}_{n}$ and $Z^{N}_{n}$ denote the wave function renormalization factor for the superconducting state ($S$) and the normal state ($N$), respectively.

%
%%%%%%%%%%%%%%%%%%%%%%%%%%%(Rysunek.4.)
%%%%%%%%%%%%%%%%%%%%%%%%%%%
%

%
%%%%%%%%%%%%%%%%%%%%%%%%%%%(Rysunek.5.)
\begin{figure}[ht]
\includegraphics*[width=\columnwidth]{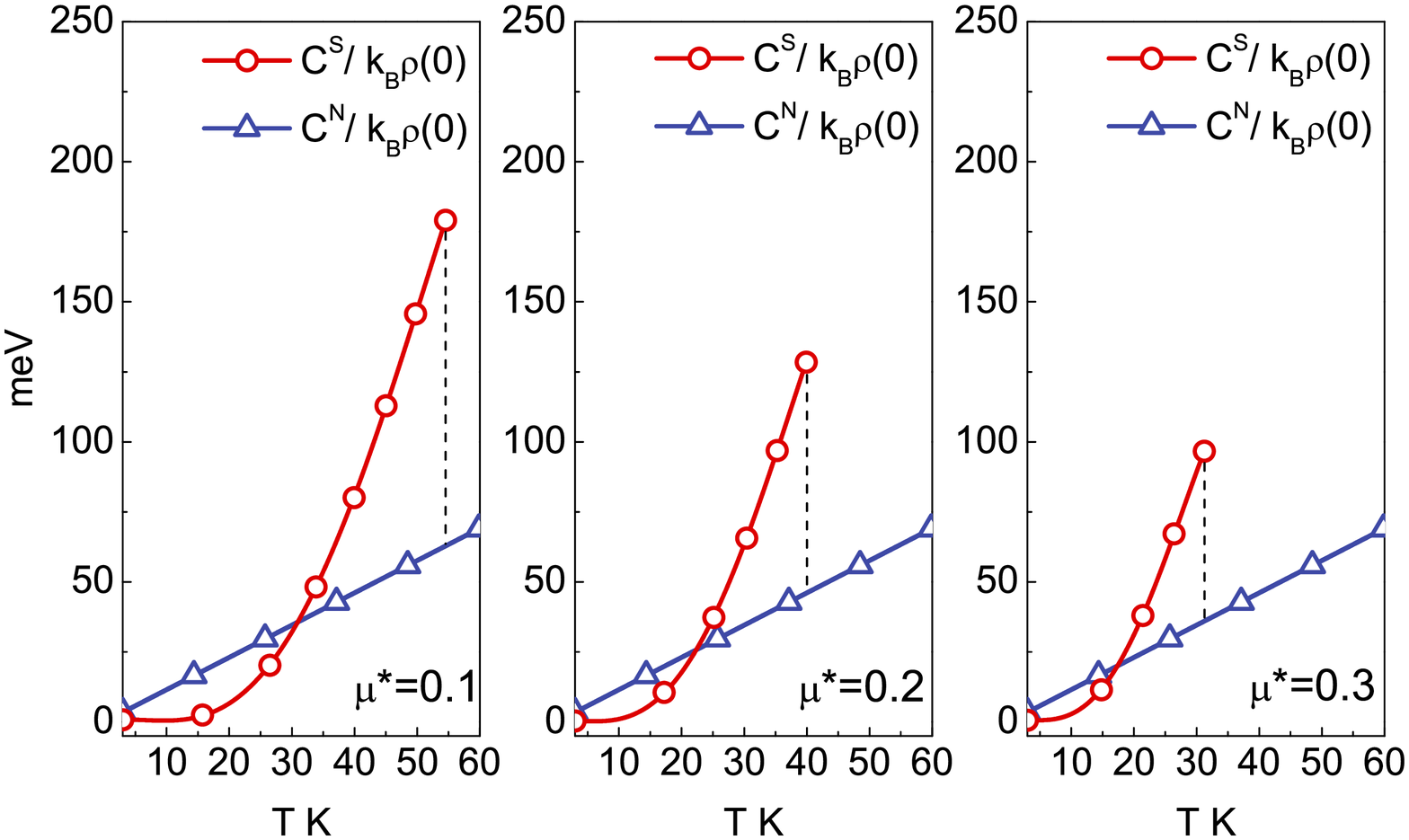}
\caption{The specific heat of the superconducting state and the specific heat of the normal state as the functions of the temperature for the selected values of the Coulomb pseudopotential.} 
\label{f5}
\end{figure}
%%%%%%%%%%%%%%%%%%%%%%%%%%%
%
In \fig{f4}, below the $0$ value, there has been presented the dependence of the free energy difference on the temperature for the selected values of the Coulomb pseudopotential. It can be easily seen that with the increase of $\mu^{\star}$ the absolute values of the free energy strongly decrease. In particular, the ratio $\left[\Delta F\left(T_{0}\right)\right]_{\mu^{\star}=0.1}/\left[\Delta F\left(T_{0}\right)\right]_{\mu^{\star}=0.3}$ is equal to $3.20$.
     
The thermodynamic critical field has been calculated by the means of the formula:
\begin{equation}
\label{r6}
\frac{H_{C}}{\sqrt{\rho\left(0\right)}}=\sqrt{-8\pi\left[\Delta F/\rho\left(0\right)\right]}.
\end{equation}
Above the $0$ value in \fig{f4} obtained course have been plotted.
Also in this case, the depairing electron correlations significantly reduce the values of the function under consideration: $\left[H_{C}\left(0\right)\right]_{\mu^{\star}=0.1}/\left[H_{C}\left(0\right)\right]_{\mu^{\star}=0.3}=1.77$, where the designation $H_{C}\left(0\right)\equiv H_{C}\left(T_{0}\right)$ has been introduced.

The specific heat difference ($\Delta C\equiv C^{S}-C^{N}$) has been estimated on the basis of the formula:
\begin{equation}
\label{r7}
\frac{\Delta C\left(T\right)}{k_{B}\rho\left(0\right)}=-\frac{1}{\beta}\frac{d^{2}\left[\Delta F/\rho\left(0\right)\right]}{d\left(k_{B}T\right)^{2}}.
\end{equation}

On the other hand, the specific heat of the normal state may be calculated by using the formula: $\frac{C^{N}\left(T\right)}{ k_{B}\rho\left(0\right)}=\frac{\gamma}{\beta}$, where the Sommerfeld constant is derived from: $\gamma\equiv\frac{2}{3}\pi^{2}\left(1+\lambda\right)$. 

The results obtained for $C^{S}$ and $C^{N}$ have been presented in \fig{f5}. It can be seen that at higher temperatures the specific heat of the superconducting state far exceeds the value of $C^{N}$, which results in the jump at the critical temperature. In addition, we can point out that the specific heat jump strongly depends on the assumed value of the Coulomb pseudopotential: 
$\left[\Delta C\left(T_{C}\right)\right]_{\mu^{\star}=0.1}/\left[\Delta C\left(T_{C}\right)\right]_{\mu^{\star}=0.3}=1.92$.

The considered thermodynamic functions allow us to calculate the values of two dimensionless ratios. 

\begin{equation}
\label{r8}
R_{H}\equiv\frac{T_{C}C^{N}\left(T_{C}\right)}{H_{C}^{2}\left(0\right)},
\qquad {\rm and} \qquad
R_{C}\equiv\frac{\Delta C\left(T_{C}\right)}{C^{N}\left(T_{C}\right)}.
\end{equation}

In the framework of the BCS theory, their values are universal and respectively equal to: $0.168$ and $1.43$ \cite{BCS1}, \cite{BCS2}. For ${\rm GeH_{4}}$ compound, it has been obtained: $R_{H}\in\left<0.147, 0.155\right>$ and $R_{C}\in\left<1.85, 1.69\right>$, when $\mu^{\star}\in\left<0.1, 0.3\right>$. On this basis, it is clear that the thermodynamic parameters of the superconducting state in ${\rm GeH_{4}}$ significantly deviate from the expectations of the BCS model.

The physical value of the energy gap at the Fermi level should be determined on the basis of the expression:
\begin{equation}
\label{r9}
\Delta\left(T\right)={\rm Re}\left[\Delta\left(\omega=\Delta\left(T\right)\right)\right].
\end{equation}

\begin{center}
\begin{figure}[ht]
\includegraphics*[scale=0.65]{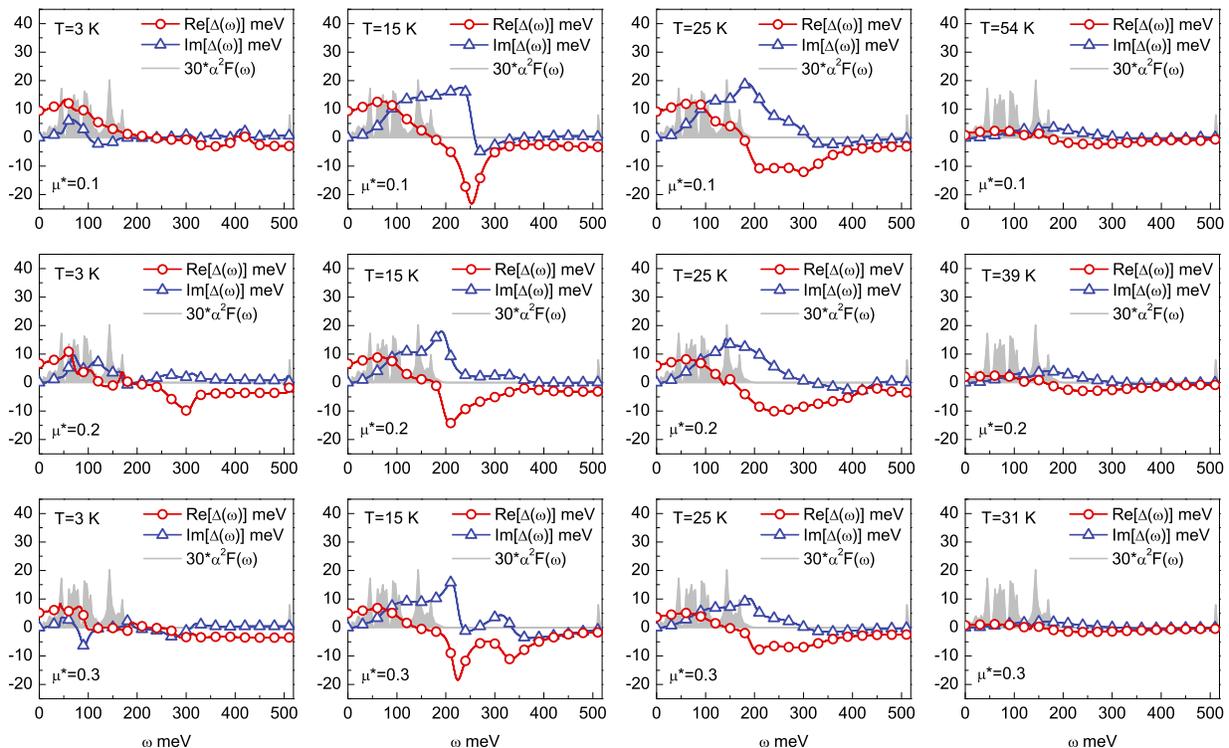}
\caption{The real and the imaginary part of the order parameter on the real axis for the selected values of the temperature and the Coulomb pseudopotential. Additionally, the rescaled Eliashberg function has been plotted.} 
\label{f6}
\end{figure}
\end{center}

The courses of the order parameter function on the real axis have been obtained by solving the Eliashberg equations in the mixed representation. The results have been presented in \fig{f6}. 

%
%%%%%%%%%%%%%%%%%%%%%%%%%%%(Rysunek.6.)
\onecolumngrid
\twocolumngrid
%%%%%%%%%%%%%%%%%%%%%%%%%%%
%

It can be seen that the order parameter takes complex values. For the lower frequencies, in the area of the applicability of the equation \eq{r9}, it can be conducted that: ${\rm Re}\left[\Delta\left(\omega\right)\right]\gg{\rm Im}\left[\Delta\left(\omega\right)\right]$. 
From the physical point of view, this result indicates the existence of the weak damping effects \cite{Varelogiannis}. 
In addition, we can note that the complex behavior of $\Delta\left(\omega\right)$, in the frequency range from $0$ to about $300$ meV, is explicitly related to the shape of the Eliashberg function. Above $300$ meV the order parameter becomes saturated.  

On the basis of \fig{f6}, we can also trace the influence of the temperature on the course of $\Delta\left(\omega\right)$. It has been found that together with the increase of $T$, the outline of the order parameter function becomes more regular, regardless of the value assumed for the Coulomb pseudopotential.

The presented results allowed us to calculate the value of the energy gap at the Fermi level for the lowest considered temperature ($2\Delta\left(0\right)$). Then, the dimensionless ratio has been estimated: $R_{\Delta}\equiv 2\Delta\left(0\right)/k_{B}T_{C}$. In the case of the ${\rm GeH_{4}}$ compound, the following result has been obtained: $R_{\Delta}\in\left<4.07,3.88\right>$, for $\mu^{\star}\in\left<0.1, 0.3\right>$. Let us notice that the BCS theory predicts that: $R_{\Delta}=3.53$. Therefore, the parameter $R_{\Delta}$ for ${\rm GeH_{4}}$ clearly exceeds the value estimated in the framework of the BCS model.

To summarize, we have calculated the values of the basic thermodynamic parameters of the superconducting state in the ${\rm GeH_4}$ compound. The calculations have been performed for a wide range of the Coulomb pseudopotential: $\mu^{\star}\in\left<0.1,0.3\right>$. It has been found that even for the large $\mu^{\star}$, the critical temperature reaches a relatively high value (see \tab{t1}). The values of the dimensionless thermodynamic ratios ($R_{\Delta}$, $R_{C}$ and $R_{H}$) significantly differ from the predictions of the BCS theory. However, with the increasing value of the Coulomb pseudopotential, the difference between the Eliashberg and BCS model fades (\tab{t1}).

It should be also noted that the value of the specific heat jump at the critical temperature and the low-temperature value of the thermodynamic critical field decreases strongly with the increasing of $\mu^{\star}$, which has been explicitly presented in \tab{t1}.

\begin{table*}
\caption{\label{t1} The values of the thermodynamic parameters of the superconducting state in the ${\rm GeH_4}$ compound for $p=20$ GPa.}
\begin{ruledtabular}
\begin{tabular}{c c c c c}
&$T_{C}$ (K) & $R_{\Delta}$, $R_{C}$, $R_{H}$ & $\Delta C\left(T_{C}\right)/k_{B}\rho\left(0\right)$ (meV) & $H_{C}\left(0\right)/ \sqrt{\rho\left(0\right)}$ (meV)\\
\hline
$\mu^{\star}=0.1$ & $54.5$ & $4.07$, $1.85$, $0.147$ & $116.31$ & $44.80$\\
$\mu^{\star}=0.2$ & $39.9$ & $3.95$, $1.79$, $0.151$ & $82.45$  & $32.30$\\
$\mu^{\star}=0.3$ & $31.2$ & $3.88$, $1.69$, $0.155$ & $60.72$  & $25.03$\\
\end{tabular}
\end{ruledtabular}
\end{table*}

%%%%%%%%%%%%%%%%%%%%%%%%%%%%%%%%%%%%%%%%%%%%%%%%%%%%%%%%%%%%%%%%%%%%%%%%%%%%%%%%%%%%%%%%%%%%%%%%%%%%

\vspace*{0.25cm}

{\bf Acknowledgments}

\vspace*{0.25cm}

This paper is dedicated to the memory of {\bf Marian Szcz{\c{e}}{\'s}niak} the member of our research group and the devoted enthusiast of the physical sciences, who passed on October 20, 2012.

The authors would like to thank Prof. K. Dzili{\' n}ski and Prof. Z. B{\c a}k for providing excellent working conditions and financial support. D. Szcz{\c{e}}{\'s}niak acknowledges also financial support by the Dean of the Faculty of Mathematics and Science JDU (grant no. DSM/WMP/11/2012/26/).

Additionally, we are grateful to the Cz{\c{e}}stochowa University of Technology - MSK
CzestMAN for granting access to the computing infrastructure built in the project No. POIG.02.03.00-00-028/08 "PLATON - Science Services Platform".

%
%%%%%%%%%%%%%%%%%%%%%%%%%%%%%%%%%%%%%%%%%%%%%%%%%%%%%%%%%%%%%%%%%%%%%%%%%%%%%%%%%%%%%%%%%%%%%%%%%%%%

%%%%%%%%%%%%%%%%%%%%%%%%%%%%%%%%%%%%%%%%%%%%%%%%%%%%%%%%%%%%%%%%%%%%%%%%%%%%%%%%%%%%%%%%%%%%%%%%%%%
%
\end{document}